\documentclass[twocolumn,showpacs,preprintnumbers,amsmath,amssymb,prb]{revtex4}
\usepackage[dvips]{color,graphics,epsfig,rotating}
\usepackage{graphicx}
\usepackage{dcolumn}
\usepackage{bm}

\begin{document}

Phys. Rev. Lett. {\bf 100}, 237003 (2008)

\title{LaFeAsO$_{1-x}$F$_x$: A low carrier density superconductor near
itinerant magnetism}

\author{D.J. Singh}
\affiliation{Materials Science and Technology Division,
Oak Ridge National Laboratory, Oak Ridge, Tennessee 37831-6114} 

\author{M.-H. Du}
\affiliation{Materials Science and Technology Division,
Oak Ridge National Laboratory, Oak Ridge, Tennessee 37831-6114} 

\date{\today}

\begin{abstract}
Density functional studies of
26K superconducting LaFeAs(O,F) are reported.
We find a low carrier density,
high density of states, $N(E_F)$ and modest
phonon frequencies relative to $T_c$.
The high $N(E_F)$ leads to proximity to
itinerant magnetism, with
competing ferromagnetic and antiferromagnetic fluctuations and
the balance between these controlled by doping level.
Thus
LaFeAs(O,F) is in a unique class of high $T_c$ superconductors:
high $N(E_F)$ ionic metals near magnetism.

\end{abstract}


\maketitle

Understanding the interplay between superconductivity and spin fluctuations
especially near magnetic phases is an important
challenge.
The discovery of
a new family of layered superconductors containing the magnetic
elements Fe and Ni and critical temperatures
up to $T_c$=26K,
specifically
LaOFeP ($T_c$=4K, and $\sim$7K with
F doping), \cite{kamihara-p,liang}
LaONiP ($T_c$=3K), \cite{watanabe}
and F doped LaOFeAs with $T_c$=26K
(Ref. \onlinecite{kamihara})
raises questions about the relationship between magnetism and
superconductivity, the origin of the
remarkably high $T_c$, and the chemical and structural
parameters that can be used to tune the properties.
Here we show that LaOFeAs is in fact close to magnetism,
with competing ferromagnetic and antiferromagnetic fluctuations,
with the balance controlled by doping.
Furthermore, we identify La(O,F)FeAs as a nearly 2D, low carrier density
metal, with modest phonon frequencies relative to $T_c$,
and high density of states.

The crystal structure is layered
with apparently distinct LaO and transition metal pnictide layers.
(see e.g. Ref. \onlinecite{kamihara})
Importantly, it forms with a wide range of
rare earths and pnictogens, with the transition
elements Mn, Fe, Co, Ni, Cu and Ru, and additionally,
related fluorides and chalcogenides are known.
\cite{kamihara-p,liang,watanabe,kamihara,quebe}
It may be that there are other superconducting compositions that
remain to be discovered in this family.

Our calculations were done
using the local spin density approximation (L(S)DA) and the
generalized gradient approximation (GGA) of
Perdew, Burke and Ernzerhof (PBE). \cite{pbe}
We used the linearized augmented planewave (LAPW)
method with tested basis sets and zone samplings for the electronic structure.
\cite{singh-book,basis}
Two independent LAPW codes were used - an in-house code,
and the WIEN2K package. \cite{wien2k}
Transport properties were calculated using
BoltzTraP. \cite{boltztrap}
The phonon dispersions were obtained in linear response
via the quantum espresso code and ultrasoft pseudopotentials
with a cutoff energy of 50 Ry and the PBE GGA.
The codes were cross-checked by comparing band structures and
zone center Raman frequencies.
We used the experimental
lattice parameters $a$=4.03552\AA, $c$=8.7393\AA, for LaFeAsO.
The internal coordinates were determined by LDA energy
minimization as $z_{\rm La}$=0.1418 and $z_{\rm As}$=0.6326.
The corresponding Raman modes
have mixed character and frequencies
of 185 cm$^{-1}$ and 205 cm$^{-1}$.
The corresponding ultrasoft pseudopotential values
were 181 cm$^{-1}$ and 203 cm$^{-1}$.
The FeAs layers consist of a square lattice sheet
of Fe coordinated by As above and below the plane to form
face sharing FeAs$_4$ tetrahedra. These are squeezed along
$c$ (the As-Fe-As angles are
120.2$^\circ$ and 104.4$^\circ$). The Fe-As distance is 2.327\AA.
The Fe-Fe distance is 2.854\AA, which is short enough
for direct Fe -- Fe hopping to be important, while the As -- As
distances are 3.677\AA, across the Fe layer, and $a$=4.036\AA, in plane.

\begin{figure}
\includegraphics[width=3.1in,angle=0]{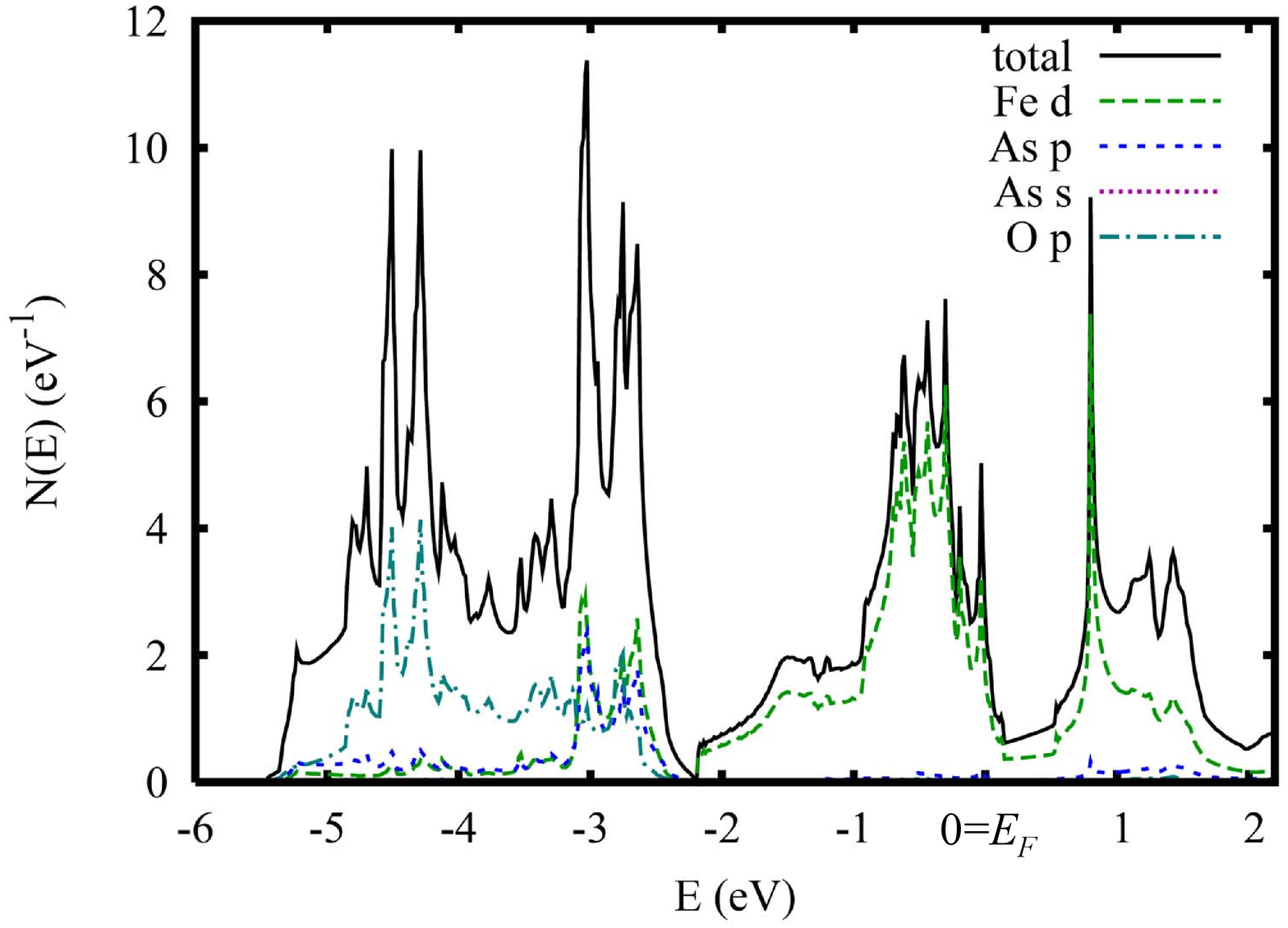}
\caption{\label{lda-dos}
(color online)
LDA density of states and projections onto the LAPW spheres
on a per formula unit both spins basis.
Note that much of the As $p$ character
will be outside the As sphere, reducing their apparent weight.
}
\end{figure}

\begin{figure}
\includegraphics[width=3.1in,angle=0]{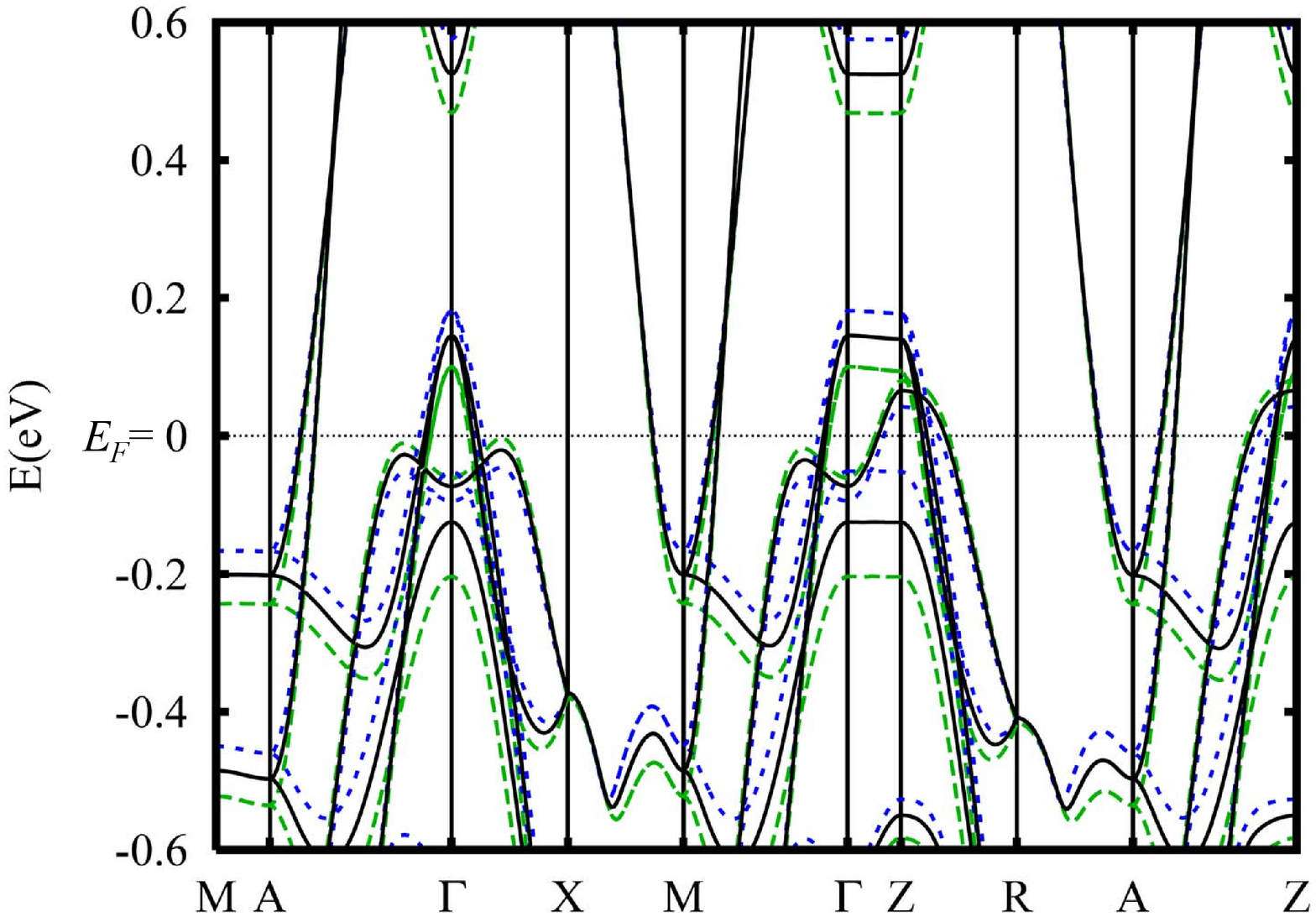}
\caption{\label{lda-bands} (color online)
Band structure of LaFeAsO
around $E_F$ showing the effect As breathing
along $z$ by $\delta z_{\rm As}$=0.04 (0.035\AA).
The unshifted band structure is indicated by the solid black line,
while the shift away (towards) the Fe is indicated by the blue dotted
(green dashed) lines.
}
\end{figure}

The electronic density of states (DOS) and band structure are
shown in Figs. \ref{lda-dos} and \ref{lda-bands}.
The band structure is similar to that
reported for LaFePO. \cite{lebegue}
Near the Fermi energy, $E_F$ it is
well described as coming from 2D metallic sheets of Fe$^{2+}$ ions
in an ionic matrix formed by the other atoms.
There is a group of 12 bands between
-5.5 eV (relative to $E_F$) and -2.1 eV. These come from
O $p$ and As $p$ states, with the As $p$ contribution
concentrated above -3.2 eV. 
These As derived bands are hybridized with the Fe $d$ states.
The Fe $d$ states account for the bands between -2.2 eV and 2 eV, with La
derived states occurring at higher energy. This is the
ordering expected from the Pauling electronegativities:
3.44 for O, 2.18 for As, 1.83 for Fe and 1.10 for La.
While there is some hybridization between Fe and As,
it is not strong and is comparable to oxides.
Thus a separation of the structure into independent LaO and FeAs
subunits is not justified from the point of view of electronic
structure or bonding and so we write the chemical formulae
in order of electronegativity.
This ionic view is supported the similarity of the Fe $d$ derived bands
of LaFePO and LaFeAsO in spite of the chemical differences between P and As.
This suggests that doping mechanisms other than replacement of O by F
may be effective, i.e. on the La site (e.g. by Th)
or the As site (e.g. by Te or Se). This may allow stabilization of the
compound with a wider range of doping levels than can be achieved using
the O site, perhaps leading to higher $T_c$.

Fe $d$ states would normally split into
a lower lying $e_g$ manifold and higher lying $t_{2g}$ states in an As
tetrahedral crystal field.
The gap between these would be at a $d$ electron count of 4 per Fe.
However, this crystal field competes with the direct Fe-Fe
interaction to yield a more complicated band structure,
with no clear gap at 4 electrons per Fe. Rather, the main feature
is a pseudogap at an electron count of 6.
The Fermi energy for $d^6$ Fe$^{2+}$ is at the pseudogap.
While we do find sensitivity of the bands near $E_F$
to the As height as shown in Fig. \ref{lda-bands},
the PBE band structure with the PBE relaxed structure is very similar
to the LDA band structure, including the details of the Fermi surface.

\begin{figure}
\vspace{0.2cm}
\includegraphics[width=3.0in,angle=0]{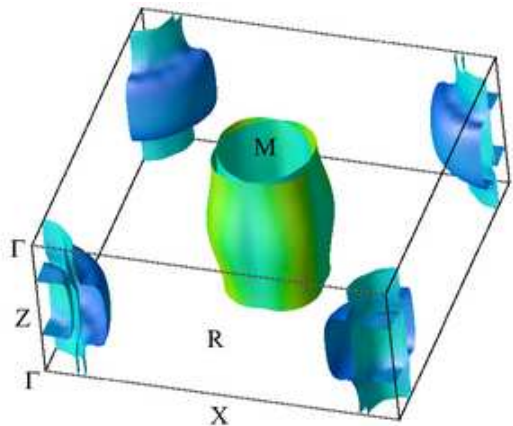}
\caption{\label{fermi}
(color online)
LDA Fermi surface of LaFeAsO shaded by velocity (blue is low velocity).
The symmetry points are $\Gamma$=(0,0,0), $Z$=(0,0,1/2),
$X$=(1/2,0,0),$R$=(1/2,0,1/2),$M$=(1/2,1/2,0),$A$=(1/2,1/2,1/2).
}
\end{figure}

The Fermi surface (Fig. \ref{fermi}).
has five sheets: two high velocity electron cylinders
around the zone edge $M$-$A$ line, two lower velocity hole
cylinders around the zone center, and an additional
heavy 3-D hole pocket, which intersects and anti-crosses with the hole
cylinders, and is centered at $Z$.
The heavy 3-D pocket is derived from Fe $d_z$ states, which hybridize
sufficiently with As $p$ and La orbitals to yield a 3D pocket.
The remaining sheets of Fermi surface are nearly 2D.
The electron cylinders are associated with in-plane Fe $d$ orbitals and
have higher velocity and will make
the larger contribution to the in-plane electrical conductivity,
The Seebeck coefficients are positive because of the proximity
to band edges in the hole bands:
$S_{xx}$=$S_{yy}$=6.8 $\mu$V/K and $S_{zz}$=8.5 $\mu$V/K, at 300K.
However, for $N(E_F)$ and other quantities that depend on the density
of states, such as spin-fluctuations, and electron-phonon coupling,
the heavier hole pockets may be more important.
Specifically, the three hole sheets together contribute 80\% of $N(E_F)$
but only 31\% of $N(E_F)v_x^2$.
The average Fermi velocities are
0.81x10$^7$ cm/s (in-plane) and 0.34x10$^7$ cm/s ($c$-axis)
for the hole sections
and 
2.39x10$^7$ cm/s (in-plane) and 0.35x10$^7$ cm/s ($c$-axis) for the electron
sections.
Including all sheets,
$v_{xx}$=$v_{yy}$=1.30x10$^{7}$ cm/s
and $v_{zz}$=0.34x10$^7$ cm/s.
This yields a resistivity anisotropy of $\sim$15 for isotropic scattering.
We note that the DOS is rapidly changing near $E_F$ and therefore
these quantities will be quite sensitive to the electron filling,
structure and other details.

The volume enclosed by the two electron cylinders (equal to
that enclosed by the hole sections) corresponds
to 0.26 electrons per cell (0.13 per formula unit).
$E_F$ lies just above a peak in the DOS, which
leads to a rapidly decreasing DOS with energy.
This peak is associated with a van Hove singularity
from the 3D hole pocket, which becomes cylindrical as $E_F$ is lowered.
The calculated value at $E_F$
is $N(E_F)$=2.62 eV$^{-1}$ per formula unit both spins.
The corresponding bare susceptibility and specific heat coefficient
are $\chi_0$=8.5x10$^{-5}$ emu/mol and $\gamma_0$=6.5 mJ/mol K$^2$
Thus LaFeAsO is a low carrier concentration, high density of states
superconductor. This is in contrast to the cuprates, which have high
carrier concentration (near half filling with large
Fermi surfaces) \cite{pickett-fs}
and lower density of states.
Recent experimental data also indicates low carrier concentration.
\cite{chen,yang,jin}
Electron doping mainly shrinks the hole pockets,
especially the 3D sheet, since these have higher mass than the electron
surfaces. This leads to an increasingly 2D Fermi surface.
Within Stoner theory a ferromagnetic instability occurs
when $NI > 1$, with $N$ now on a per spin basis. Since the
DOS near $E_F$ is close to pure Fe $d$ in character it is appropriate
to choose $I$ $\sim$ 0.7 - 0.8 eV, which would put LaFeAsO very close
to a magnetic instability.

\begin{figure}
\includegraphics[height=3.2in,angle=270]{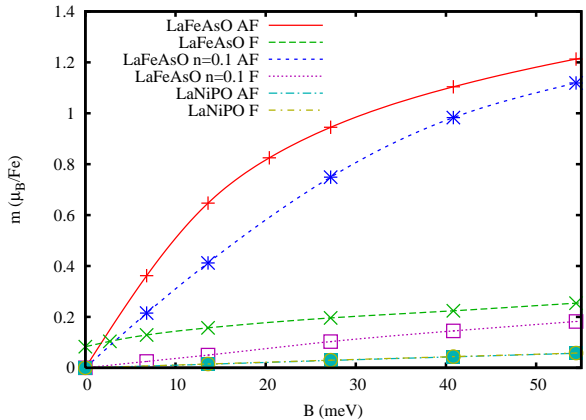}
\caption{\label{ind} (color online)
Induced moments as a function of antiferromagnetic (AF) and 
ferromagnetic (F) field.
The field is applied by an extra spin dependent potential
+/- $B$ inside the Fe/Ni LAPW spheres.
}
\end{figure}

Fixed spin moment calculations were done
for the LaFeAsO, LaFePO and LaNiPO, and for
LaFeAsO with
O replaced by virtual atoms, $Z$=7.9 and $Z$=8.1 (to simulate
doping). \cite{virt}
In the LDA LaFeAsO is indeed on the
borderline of a ferromagnetic instability and electron doping
moves away from this instability.
The ground state is an itinerant ferromagnet
with a moment of 0.08 $\mu_B$/Fe.
However, the small energies involved
are below the precision of the calculation,
so it may only be concluded that the material is on the
borderline of ferromagnetism.
This borderline behavior, where the energy is independent	
of magnetization, persists up to $\sim$ 0.2 $\mu_B$/Fe. This is
close to the value where the holes become fully polarized.
This places stoichiometric LaFeAsO near a ferromagnetic
quantum critical point.

The experimental susceptibility, \cite{kamihara} $\chi(T)$ of undoped LaFeAsO
is weakly temperature dependent between 20K and 300K, with value
$\sim$ 50x10$^{-5}$ emu/mol,
and increases strongly at lower $T$. While the low $T$ $\chi(0)$ is
not known, even taking the higher temperature value, the implied Stoner
renormalization $(1-NI)^{-1}$ from $\chi_{exp}/\chi_0$ is 6 and is probably
significantly higher depending on the observed upturn below 20K.
By comparison $\chi_{exp}/\chi$ is $\sim$5 for MgCNi$_3$, \cite{singh-mgcni3}
$\sim$7 for Sr$_2$RuO$_4$, \cite{maeno} and $\sim$9 for Pd metal.
Ferromagnetic spin fluctuations are strongly pair breaking for singlet
superconductivity both in the $s$ channel favored by electron-phonon
interactions, and in $d$ channels, such as in the high-$T_c$ cuprates.
Thus such fluctuations would suppress $T_c$.
However, it is conceivable
that a sufficiently strong electron phonon interaction,
which is needed to explain the $T_c$ $\sim$ 26K in that
scenario, could overcome the pair breaking effects of ferromagnetic
fluctuations as was discussed for MgCNi$_3$.
\cite{singh-mgcni3,ignatov}
Turning to the F doped material, which is the actual superconducting phase,
$\chi_{exp}$ is higher than for the stoichiometric material
and increases as $T$ is reduced over the whole temperature range,
rising to 
$\sim$ 200x10$^{-5}$ emu/mol at $T_c$.
This is opposite to the trends with band filling in $N(E_F)$
and the fixed spin moment curves.
The most simple explanation would be in terms of secondary phases.
Another explanation is that the undoped compound is magnetically ordered.
There is in fact a resistivity peak at $\sim$ 150K and
a minimum at $\sim$ 100K, \cite{kamihara}
but $\chi(T)$ does not show strong changes at these temperatures.

We did calculations with
an antiferromagnetic arrangement of the Fe spins in the square lattice.
As for ferromagnetic alignment, we find
a borderline instability,
with stable moments in the virtual crystal
calculations at lower electron count while the proximity
to antiferromagnetism is reduced with increasing electron count.
We also did calculations with small fields
applied to the Fe LAPW spheres in both ferromagnetic and antiferromagnetic
configurations. The resulting moments, defined by the magnitude of the
spin polarization in each Fe LAPW sphere, are shown in Fig. \ref{ind},
along with results of similar calculations for LaNiPO and LaFePO.

\begin{figure}
\vspace{0.2cm}
\includegraphics[width=3.2in,angle=0]{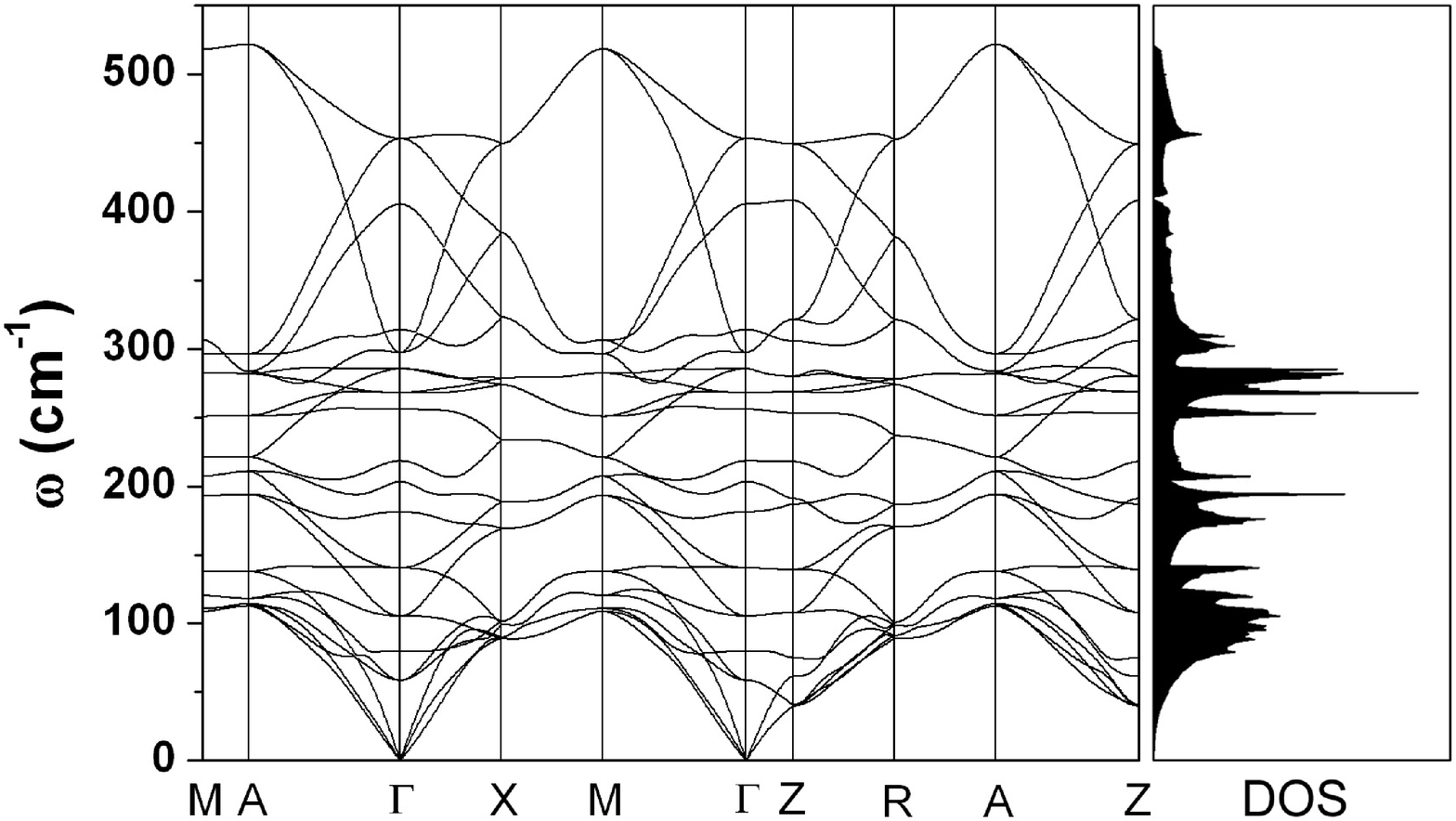}
\caption{\label{phonons}
Phonon dispersions and density of states of LaFeAsO.
}
\end{figure}

While ferromagnetism is initially favored, the antiferromagnetic
response becomes larger for higher fields with a cross-over at an
energy scale of 30K. Thus thermal fluctuations at higher temperature
and perhaps quantum fluctuations would both favor antiferromagnetic
fluctuations over ferromagnetism.
This nearest neighbor antiferromagnetism is related to superexchange.
\cite{sdw}
Doping with electrons
rapidly reduces the proximity to ferromagnetism.
Furthermore, it should
be emphasized that LaNiPO, which is also a superconductor
has much lower susceptibility for both
ferromagnetic and antiferromagnetic fluctuations.

The calculated phonon dispersions and density of states of LaFeAsO
are shown in Fig. \ref{phonons}.
The acoustic modes are not strongly anisotropic.
In particular, there are no soft elastic constants associated with 
shearing of the planes along the $c$-axis.
The Debye temperature from the acoustic mode velocities
is $\Theta_D$=340K with an uncertainty of 10\% due to sampling.
Turning to the optic modes, the dispersions may be divided into two
regions. Below $\sim$ 300 cm$^{-1}$ the dispersions are dominated by
metal and As modes of mixed character,
while the modes above $\sim$ 300 cm$^{-1}$ are strongly O derived.
The phonon density of states shows three main peaks below 300 cm$^{-1}$.
These are all of mixed metal and As character, and are
a large peak centered at $\sim$ 100 cm$^{-1}$ and smaller peaks at
$\sim$ 170 cm$^{-1}$ and $\sim$ 280 cm$^{-1}$.

The phosphides LaFePO and LaNiPO also superconduct,
though with lower $T_c$. The band structures near
$E_F$ of LaFeAsO and LaFePO (Ref. \onlinecite{lebegue}) are very similar.
We find that LaFePO, like LaFeAsO is on the borderline of a 
ferromagnetic state in the LSDA.
As expected from the different electron count, the electronic
structure of LaNiPO is very different. Calculations
yield large two dimensional Fermi surfaces and
lower $N(E_F)$.
With the assumption that the origin of superconductivity is the same
in the Fe and Ni compounds,
spin fluctuation mechanisms seem unlikely since
these depend on a match between the ${\bf q}$ dependent
susceptibility and the Fermi surface, requiring different
coincidences for
the Fe and Ni compounds.
One scenario would be that
there are strong electron phonon interactions related to the
layered crystal structures and the presence of high charge ionic species
in proximity to metallic layers.
Then ferromagnetic spin
fluctuations are pair breaking and the main difference between the
P and As compounds is the heavier mass of As, yielding a trend
opposite to the usual isotope effect.
Conversly, if the superconductivity of the
Ni compound has a different origin (note
the low $T_c$ of LaNiPO) a common spin-fluctuation based mechanism for
the two Fe compounds is likely. Based on the suppression
in the ferromagnetic susceptibility on doping the relevant fluctuations
would be antiferromagnetic in nature.
In both scenarios, doping plays a key role in the Fe based
compounds. Electron doping reduces $N(E_F)$,
which would lower pairing strength by reducing the phase space, and
at the same time strongly suppresses ferromagnetic fluctuations,
which are strongly pair breaking for singlet superconductivity.
Furthermore, antiferromagnetic spin fluctuations are also weakened
by electron doping, which is of relevance if these are involved in
the pairing.
One way of distinguishing these pictures would be via doping on the Fe site,
e.g. with Co or Ni. Since the electronic structure near $E_F$ is from
Fe $d$ bands, such doping would be much more strongly
scattering than doping in the O layer and would be very
detrimental to non-$s$ wave spin fluctuation mediated superconductivity
but not to $s$-wave superconductivity.
It will be interesting to map out the
similarities and differences between LaFeAsO and other
superconductors with ionic metal type electronic structures,
and systematically explore the relationship
between spin-fluctuations and superconductivity.

We are grateful for discussions with D.G. Mandrus, B.C. Sales, R. Jin,
and M. Fornari and support from DOE,
Division of Materials Sciences and Engineering.


\begin{references}
\bibitem{virt}
The virtual crystal approximation differs from the rigid band
approximation as it self-consistently includes the average
change in potential, but neglects randomness.
It is generally valid when the
scattering effect of disorder is weak, as may be expected here
because O does not participate in the states near the Fermi energy,
and the O ions are located far from the Fe planes.
\end{references}
\end{document}